\documentstyle[12pt]{article}
\topmargin - 1.5cm
\leftmargin - 1cm
\begin{document}      

NEW BOOK

\begin{center}
{\Large \bf OBEDIENT QUANTUM MECHANICS : \\ NEW STATUS OF THEORY \\ in 
inverse problem approach  (in pictures)}
\end{center}

\vspace*{3cm}

ABSTRACT

New status  in quantum mechanics is connected with  recent
achievements in the inverse problem. With its help instead of about ten
exactly solvable models which serve as a basis of the contemporary
education there are infinite (!) number, even {\bf complete} sets of such
models. So, the whole quantum mechanics is embraced by them. They 
correspond to all possible variations of spectral parameters which 
determine all properties of quantum systems.  There appears a possibility 
to change at wish quantum objects by variation of these parameters as 
control levers and examine quantum systems in different thinkable 
situations. As a result, we acquire a vision of the intrinsic logic of 
behavior of any thinkable system, including real ones. The regularities 
revealed by computer visualization of these models were reformulated into 
unexpectedly simple universal rules of arbitrary transformations and what 
is more, their elementary constituents were discovered (new breakthrough). 
Of these elementary "bricks" it is possible in principle to construct 
objects with any given properties.  This book of inverse problem quantum 
pictures is utmost intelligible and recommended to any physicists, 
chemists,  mathematicians, biologists  from students to professors who are 
interested in laws of the microworld.
\newpage

{\Large \bf CONTENTS 
}

{ \bf PREFACE
}

{\bf INTRODUCTION
}

1 {\bf  Main notions about control of energy level positions  
 }

2 {\bf  Main notions about control of spectral weights of states
 (their localization  in space)}
     
\hspace*{0.25cm} 2.1 Removal of bound state energy levels from spectrum and 
creation of new ones

\hspace*{0.25cm} 2.2 Bound states embedded into continuum

3 {\bf Inverse problem and supersymmetry formalisms in quantum 
 mechanics}
 
\hspace*{0.25cm} 3.1 Elements of inverse problem theory

\hspace*{0.25cm} 3.2 Inverse problem on  a finite interval

\hspace*{0.25cm} 3.3 Inverse problem on the half axis: Gelfand-Levitan 
and Marchenko approaches
 
\hspace*{0.25cm} 3.4   Exactly solvable models of inverse problem

\hspace*{0.5cm} 3.4.1 Transformations of the discrete spectrum

\hspace*{0.5cm} 3.4.2 Transformations of the continuous spectrum

\hspace*{0.25cm} 3.5 Supersymmetry transformations 

\hspace*{0.5cm} 3.5.1 Darboux transformations of Schroedinger operator 

\hspace*{0.5cm} 3.5.2 Creation of energy level in oscillator potential 
and chains of   transformations

\hspace*{0.5cm} 3.5.3  Multistep supersymmetry 
transformations and energy level  shifts

\hspace*{0.5cm} 3.5.4 Two spectra theorem and supersymmetry transformations

\hspace*{0.5cm} 3.5.5 Supersymmetry and inverse problem

\hspace*{0.25cm} 3.6 Approximate solutions

\hspace*{0.5cm} 3.6.1 Convergence of approximations and stability of solutions

\hspace*{0.5cm} 3.6.2 Influence of the upper part of spectrum on the shape of 
potential

\hspace*{0.5cm} 3.6.3 Reproducing (reconstruction) of potentials by multiple
 solution of   the  direct problem
 
 4 {\bf "Addition" of elementary transformations  }

\hspace*{0.25cm} 4.1 Approximation by exact models from 
their complete set

\hspace*{0.25cm} 4.2 Some instructive combinations of shifts over E 
and x

 5 {\bf "Annihilation"  of states  by degeneration of energy levels}

 6. {\bf Elements of resonance theory. Tunneling, packets and periodical 
 potentials}

\hspace*{0.25cm} 6.1 Target with delta barrier

\hspace*{0.25cm} 6.2 Resonance control (in case of potentials of finite 
range on half axis)

\hspace*{0.5cm} 6.2.1  Degeneration of resonances

\hspace*{0.25cm} 6.3 Levinson theorem

\hspace*{0.25cm} 6.4 The total quantum reflection at chosen energy values

\hspace*{0.5cm} 6.4.1 New aspects of resonance transparency

\hspace*{0.5cm} 6.4.2 Resonance tunneling control in inverse problem approach

\hspace*{0.5cm} 6.4.3 Periodical structures: elements of spectral zones control

\hspace*{0.25cm} 6.5 Unusual (non-Gamov) decaying states (energy shifts 
into complex plane)

\hspace*{0.5cm} 6.5.1 Transformation of bound states
 
\hspace*{0.5cm} 6.5.2 Transformations os scattering states

 7 {\bf Wave mechanics on lattices}

\hspace*{0.25cm} 7.1 Elements of zones theory (one allowed zone)

\hspace*{0.5cm} 7.1.1 Free waves of finite difference Schroedinger equation

\hspace*{0.25cm} 7.2 Peculiarities of wave motion over lattices with 
piecewise constant potentials

\hspace*{0.5cm} 7.2.1 Tunneling (new aspects)
 
\hspace*{0.25cm} 7.3 Resonances on lattices

\hspace*{0.5cm} 7.3.1 Tunnelling resonance in allowed zone

\hspace*{0.25cm} 7.4 Waves on lattices  with linear potential  V(n) = C n

\hspace*{0.5cm} 7.4.1 Bessel functions as exact solutions of the difference 
Schroedinger  equation

\hspace*{0.25cm} 7.5 Reconstruction of spectra of systems on lattices

\hspace*{0.5cm} 7.5.1 Energy shifts $E_{n}$

\hspace*{0.5cm} 7.5.2 Equidistant spectrum and spectrum with gap

\hspace*{0.5cm} 7.5.3 Inversion of spectrum

\hspace*{0.5cm} 7.5.4  Reflectionless potential perturbations on lattices (SUSYQ approach)

\hspace*{0.25cm} 7.6 Equations of higher order ($\ge 4$)

\hspace*{0.5cm} 7.6.1 Models

\hspace*{0.5cm} 7.6.2 Splitting of equations of higher order 

8 {\bf Systems of coupled Schroedinger equations}

\hspace*{0.25cm} 8.1 Uncoupled channels

\hspace*{0.25cm} 8.2 Coupled channels

\hspace*{0.25cm} 8.3 Variations of spectral weight vectors

\hspace*{0.25cm} 8.4  Interchannel oscillation of packets

\hspace*{0.25cm} 8.5  Violation of "natural" law of more rapid asymptotic 
decrease of wave functions in channels in more closed channels

\hspace*{0.25cm} 8.6 Multichannel peculiarities of states degeneration

\hspace*{0.25cm} 8.7 New mechanism of complete reflection and transparency

\hspace*{0.5cm} 8.7.1 Resonance multichannel tunneling and reflection 

\hspace*{0.25cm} 8.8 Multichannel generalization of the inverse problem  
and SUSYQ formalism 

\hspace*{0.5cm} 8.8.1 Exactly solvable models (complete sets)

\hspace*{0.5cm} 8.8.2 Peculiarities of multichannel SUSYQ transformations

9 {\bf Multidimensional and many-body problems}

\hspace*{0.25cm} 9.1 Reduction to the multichannel equations

\hspace*{0.25cm} 9.2 Different thresholds of excitations of channels

\hspace*{0.5cm} 9.2.1 Sturm functions as basis in scattering problems

\hspace*{0.5cm} 9.2.2 Spheroidal systems

\hspace*{0.5cm} 9.2.3 Exactly solvable models 

\hspace*{0.25cm} 9.3 Many-body systems

\hspace*{0.5cm} 9.3.1 Three-body one-dimensional models 

\hspace*{0.5cm} 9.3.2 Boundary conditions

\hspace*{0.5cm} 9.3.3  Collective Penetration of barriers 

\hspace*{0.5cm} 9.3.4 Collective excitations    

\hspace*{0.5cm} 9.3.5 Hypersymmetrical basis, K-harmonics 

\hspace*{0.5cm} 9.3.6 States of continuum spectrum  
(discrete parametrization)

\hspace*{0.5cm} 9.3.7 Interaction regions of 4-body system 
 (constituents of the structure) 

{\bf Comments on literature}

{\bf Reference list}

{\bf Subject index}

\newpage

{\bf \Large Preface}
\protect \addcontentsline{toc}{chapter}{Preface}

\par
\parbox[h]{6cm}{.}\parbox[h]{7cm}{
\medskip
"The method of the inverse problem is without any hesitation one of the
most beautiful discoveries in  the mathematical physics  of XX century"
\par  Novikov S.P., Zakharov V.E., Manakov S.V., Pitaevsky L.P.
}

\vspace*{.5cm}
  Quantum mechanics has recently celebrated  100 years  of  the
notion of quant (M.Plank) and  75 years of the Schr\"odinger equation
which determines from the given  potential the wave function and hence
the properties of the investigated object, the {\bf direct problem}.
Only in a quarter of a century there  appeared  the inverse
problem (IP) equations of  Gelfand-Levitan-Marchenko \cite{GL,Mar}.  This
achievement of Russian mathematicians gives us the all penetrating
 $\,\,$
v i s i o n. There are no tools like our  eyes with microscopes for
examining details of such small objects as  atomic and
nuclear systems.  For this we need "illumination" with such
ultra-short waves which cannot be operated as
the visible light. It appears that the formalism of IP
replace our eyes here. This is a wonderful present of
mathematicians, which for a long time was not properly realized and used by
physicists.

Quantum mechanics like the Moon was known  only from  one side (direct
problem).  This absolutely unsatisfactory situation with quantum
education remains till now. So we
need the global program of raising the level of professional
qualification and this book is intended for the physical community
from students  to professors. With its help they will acquire the new 
knowledge with the minimal efforts.

Standing "on  the shoulders of the giants" (being based on the results of 
the fathers-creators of the IP) we can succeed with computer visualization 
of exactly solvable models (ESM) in  widening the horizon of quantum 
literacy.  So, for us the whole comprehension  of the 
 contemporary wave mechanics was significantly  transmuted.  These were 
 the fascinating lessons on quantum intuition which we would like 
 to  share with readers.

The suggested monograph is not to serve as substitution for the already
published textbooks on quantum mechanics, but as important 
addition.   The reader will be acquainted  with
the {\bf algorithms of elementary and universal transformations : the
potential perturbations which change any chosen spectral parameter keeping
 all other unperturbed}.

  So, the notion is acquired   about  {\bf the simple  "building blocks" of
which as with  children toy constructor  set can be composed
quantum systems with desired properties}.  So, it is
possible (at least theoretically) to transform gradually one quantum
system into any other one. This illustrates the methods of the
theory of spectral, scattering and decay control. This is such a degree
of quantum mastery about which it was difficult
even to dream in the direct problem approach.  And  it is done following
the  discovered in IP very simple and exact formulae of the infinite
number, even ({\bf complete sets!} of exactly solvable models (ESM), these
"milestones of the cognition". Meanwhile the most recent textbooks
 give the information about hardly  ten such models (rectangular, 
oscillator, Coulomb , etc.  potentials).

To the first successes on this way were devoted our books \cite{Les}:
"Lessons on Quantum Intuition", 1996 and  "New ABC of Quantum Mechanics in
pictures", 1997. The best results from them was included in this book, but
with completely renewed illustrations, which were selected from the
thousands we have got during the recent years.

 As  a principle of exposition  we have used the  comparisons
 of the direct and inverse problems, of peculiarities of
wave motion over the discrete lattices and  continuum variables,
of the one- and multichannel processes, of the solutions for the second and
higher order equations.
This book is based on the lecture courses \cite{Les} delivered at the
leading Moscow universities (MSU, Phys.-Tech., MEPHI) and reports in $\sim
150$ centers of the world.  

It is desired that the reading of this book will be maximally creative 
process.  Authors are particularly interested in cooperation with the 
readers in different forms. Please write us (zakharev@thsun1.jinr.ru;
$\,\,$ chabanov@thsun1.jinr.ru)
 your comments on  aspects you will like in this book and any criticism
 and your suggestions, too.
  You are invited to
our homepage: http://thsun1.jinr.ru/$\sim$zakharev/.
                                
\newpage

INTRODUCTION

\parbox[h]{3.9cm}{.}\parbox[h]{10.cm}{\par \hspace{1.cm}
 "I wanted most to give you some appreciation of the wonderful world and the
physicist's way of looking at it, which, I believe, is a major part of the
true culture of modern times. (there are probably professors of other
subjects who would object, but I believe that they are completely wrong).
Perhaps you will not only have some appreciation of this culture; it is
even possible that you may want to join in the greatest adventure that the
human mind has ever begun" Richard P.Feynman


 }

\vspace*{0.5cm}
 {\bf Infinite} examples of solving the Schr\"odinger equation
during several decades from its creation have convinced
physicists that arbitrary variations of potentials   $V$
 lead  to perturbation of almost  {\bf all} spectral characteristics
($S$).
This  witnessed also about complicated  {\it nonlinear} connections of
interactions   $V$ with observables ($S$)
in spite of linearity of the Schr\"odinger
equation itself.  It seemed that  in this situation
even a suspicion about the possibility of {\bf independent variations of
separate $S$ parameters} could not appear.
Really, till now the common reaction of an audience to information about
such delicate spectral control (including the authors of
textbooks on quantum mechanics) is used to be  expressed in emotional
words of type:  "is it possible ??!".

Such possibility to change the chosen spectral parameter with this sniper's
accuracy would be fantastically great because the spectral parameters could
be the levers of flexible and precise control of the physical properties.
This was discovered by using the inverse problem (IP)
formalism, where  $S$  are the  {\it input} parameters. Traditionally, IP 
 was considered as restoration  $S\rightarrow V$ of a {\it concrete} 
potential from experimental data $S$.  We ourselves have not understood for 
a long time that it is incomparably more interesting to use 
potential transformations $\stackrel{\circ}{V} \to V$  for construction of 
arbitrary quantum systems.

  This new status of IP is promoted by existence of exact
analytical expressions for the transformed potentials and the wave functions
({\it exactly} solvable models -- ESM) which correspond
to the changes of separate spectral parameters  $S$.
  The complete set of
$S$-parameters determines the relevant
potential {\it uniquely}.  So, the sequence of ESM  corresponding 
to variations of the picked out parameters allow one in principle to 
approximate gradually with an arbitrary precision  {\it any} 
imaginable object, see, for example, Fig.4.1.

 All these
innumerable ESM of the potential transformations  $\Delta V= V -
\stackrel{\circ}{V} $ are simple building blocks  in themselves. The 
computer visualization allowed us to clarify that they are in their turn 
combined of the simplest previously  unknown fundamental constituents. 
There were revealed corresponding elementary universal algorithms of 
constructing quantum systems.

Nothing like this we have expected to discover in quantum  mechanics.
This gives a more  deep understanding of quantum theory (its essence, the
relations    $S \rightarrow V $). The particular value of these algorithms
is their {bf qualitative} predictive power, which has already helped us to 
find new effects, for example, those listed below in this introduction.  
Without this, their search  would be like a blind hope to find  "a needle 
in a hay stack". It is often possible to succeed in 
revealing the key points of the new  effects even without computer. Such a
quantum intuition allows one to economize the efforts, memory and time. 
Here is a {\bf rare combination of the simplicity with the mathematical 
precision}.

The corresponding mathematical formalism will be also presented in a 
maximal accessible form.  The IP is introduced as generalization of the 
widely known procedure of Gramm-Schmidt orthogonalization to the case of 
infinite, and what is more, continuous "number" of vectors. The reader 
will be acquainted with the methods of supersymmetry (SUSYQ). 
The IP  ESM's turn out to be   a subset  of ESM generated by SUSYQ
This  increases  still more the possibilities of the quantum control.
For example, sometimes with SUSYQ the energy 
level shifts becomes infinitely more simple procedure.

For most physicists, the IP is still "the thing in 
itself".  So we try  to share our discoveries which 
transform the {\bf antiintuitive aspects of the theory 
into obvious ones}  with everybody who is interested in new 
quantum literacy.

We shall briefly enumerate some "wonders" of the quantum design to give a 
notion about most essential points in the book contents.  
In order to illustrate corresponding examples we shall refer
most impatient readers to selected important  figures  scattered 
throughout the book. Here, it is necessary to realize that 
each of these wonders is not only a separate point in the space of quantum 
knowledge, but a bundle of rays elucidating significant 
regularities hidden before, and all together this points  compose the 
supporting manifold of notions renovating our quantum intuition.

 $\diamondsuit$  We begin 
with {\bf shifting} arbitrary separate bound state energy level $E_{n}$ 
{\it keeping all other spectral data unchanged}. Fulfilling an infinite 
number of these additional conditions by choosing the 
proper potential in direct problem approach  was a hopeless thing. In IP 
and SUSYQ 
approaches it is one of simplest standard operations. The following  
qualitative rule was discovered for the shifts up (down) of a level. 
{\bf Each bump of the chosen state must be pushed by the 
potential barrier (well) in the vicinity of the bump extremum, and for 
conservation the positions of other levels there must be compensating wells 
(barriers) in the vicinity of the knots of the chosen state}. See the 
examples for one bump  in Figs.1.1, 1.2 and exact formulae 
(\ref{vE1down,psiE1down}).

 $\diamondsuit $ The mutual coincidence of spectral parameters in two 
different systems 
(1,2) for lowest bound states brings together the profiles of the 
corresponding  potentials $V_{1}, \enskip V_{2}$ in the proper 
 energy region.  This is a qualitative illustration of the theorem about a 
 unique correspondence between the potential and its spectral data. In a 
particular case, it can be achieved by the energy level shifts of bound 
states in the above-mentioned way,  as is shown in Figs 4.1, 4.2.
                 
 $\diamondsuit $ The same  algorithm of the level shifts in the infinite 
deep rectangular well can be applied to the {\bf control of spectral zones 
of the periodical systems}. The corresponding potential perturbation must 
be periodically continued from the given interval. As a result, the chosen 
zone of the band spectrum can be shifted over the energy scale (see, e.g., 
Fig.6.22).  Really, each energy level of a system in a separate finite 
interval is expanded into the allowed zone after the periodical 
continuation (appearance of the spectral band structure).  So, the shift of 
one such level give rise to the shift of the spectral zone, too.  We can 
join in this way the neighbor allowed zones, which means the {\bf 
liquidation of one lacunae (forbidden zone)}. 

 $\diamondsuit $ The method analogous to the energy level shift allows one 
to {\bf tear from the  continuous spectrum its  lowest state, lower 
boundary ($E=0$), transforming it into bound state }  ($E_{n}<0$), see 
Fig.1.6. So, with the initial system of a free wave motion  
($\stackrel{\circ}{V}(x)=0$)  we get {\it absolutely transparent} 
potentials. This is because  the considered elementary transformations do 
not violate the spectral properties of other states  of the continuous 
spectrum  of free  initial motion.

 $\diamondsuit $ The {\bf shift of the chosen bound state into the complex 
energy plane,
transforms the corresponding initial stationary state into the decaying 
one}.  {\it Only in contrast with the usual Gamov decaying states} these 
states have not unrestricted growth in any direction.

 $\diamondsuit $ It is possible to construct still unknown  {\bf periodical 
potentials without(!) spectral lacunae} with such  shifts $E 
\rightarrow E+i \gamma $ of the continuum spectrum states.

 $\diamondsuit $ You will also see {\bf how  to shift the localization of 
the wave function } $\Psi_{n}(x)$ of the chosen  $n$th state. It means the 
control of the probability density distribution  via 
the change  of the {\it spectral weight factor} $á_{n}$) (SWF).  
For example, the wave function can be pressed to the origin, see Fig.2.1.  
Variations of different SWF $c_{n}$ together with the shifts of energy 
levels  compose a  {\bf complete set}  of arbitrary potential 
transformations.  It unexpectedly appears  possible to understand how  the 
individual bumps of the wave functions are transformed by variation of 
a SWF and what simplest potential perturbations  are needed for that.  
{\bf The block well-barrier (or barrier-well) shifts the corresponding bump 
to the left (right).  Simultaneously, all other states undergo some 
 recoil in the opposite direction, so that there  occurs  separation 
of the chosen state}, see Figs 2.1, 2.2.  This is an  important element of 
the quantum literacy, which has never before been mentioned.

$\diamondsuit $ Remarkable  itself, the same algorithm appears to be suitable 
also for {\bf removing the given energy level or creation of a new level at 
the prescribed place (position)} of the bound states discrete spectrum, 
keeping all other energy levels untouched.
  When the SWF $c_{n}$ is changed to zero, see eqs. (\ref{psi}) and
(\ref{V}), the $n$th bound state is removed to infinity by a 
carrier-potential, see  Fig.2.5.

 Analogously, the creation of bound states can be considered as carrying 
them from "behind the horizon" to the desired place.

$\diamondsuit $ The same algorithm of shifting states to the left (right) in the
infinite rectangular well can be used for  {\bf tearing the continuous
spectrum by creation of lacuna, the forbidden spectral zone of
the periodic system} at the given place on the energy scale.
For this, it is needed to continue periodically the potential perturbation
for shifting the state with the energy $E_{n}$. Really, the symmetry 
violation of the eigenfunction $\Psi_{n}(x)$ pressed to one of the ends of 
the separate interval  results in divergent solutions for the periodical 
structure. This is because smooth matching of the unsymmetrical solutions 
on the adjacent intervals (with  periodical potential 
perturbation continuation) is possible only with exponentially increasing 
oscillations. Such behavior is characteristic of solutions in the forbidden 
zone. For the well-known analogy it can be compared with the exponential 
increase in solutions under the potential barriers. Since we can arbitrary 
change the value  of SWF $c$ for arbitrary auxiliary eigenvalue problems,  
the degree of forbiddeness in an arbitrary energy point of the 
spectrum is under our control.

  There are exact solutions corresponding to finite gap potentials 
($N<\infty$ lacunae) \cite{}. In the above-mentioned  approach induced by
finite interval IP eigensolutions with spectral parameters $E_{\nu }; \, 
c_{\nu }$ we achieved a significant extension of the ESM classes with 
$N=\infty$. 

$\diamondsuit $ {\bf If two neighbor energy levels are shifted to one another 
in symmetrical potentials $V(x)=V(-x)$, there occurs splitting  of the 
corresponding states which parts are going from one another until they 
disappear at $\pm \infty$ (effective annihilation) }. It is caused by the 
mutual incompatibility of orthogonal states, see Figs 5.1, 5.7.
  For nonsymmetric potentials there can be annihilation of 
only one of the two degenerating states.

 {\bf The phenomenon of annihilation of  the spectral
zones of periodical structures can occur also when they are shifted to one
another}, see Fig.6.22.

  It will be also shown how  {\bf to force the more deep 
quasibound state behind the potential barrier to decay more intensively 
than the upper ones} by using $c$-shifts of space localization of the wave 
functions. The example of such a decay (resonance widths) control  is shown 
in Fig.6.7.

$\diamondsuit $ It is also possible  {\bf to transform the scattering state 
into the bound state} embedded into the continuum at the same energy point  
$E$. It is done by an infinite number of the above-mentioned potential 
well-barrier blocks.  It seems that the wave with the positive energy must 
fly away from the region of initial concentration. Really, due to 
interference it appears to be confined even at the energy above the 
barriers, see Fig.2.14.

$\diamondsuit $ With the potentials confining waves at the origin on the 
half axis, we constructed  potential on the whole axis  $x$  with {\bf 
100\% reflection at the given energies $E_{\nu }$ and even above barriers} 
\cite{PRL}.  Such a resonance reflection has not yet been considered in wave
mechanics. It is significant that the parameters of these resonances (their
  widths, positions and number) are under control, see Figs 6.13, 6.14.

$\diamondsuit $ We shall give the simple original explanation of resonance 
tunneling (total transparency at $E=E_{\nu }$ under barriers). It will be 
shown how an additional flux coming from the opposite direction can help or 
hinder the tunneling of initially incident waves which is often not 
suspected (see section \ref{newasrt}).

The rules of bound state creation at  $E>0$ and $E<0$ can be generalized
to the forbidden and allowed zones of periodical structures.

The effects enumerated above and those to be
considered further are not independent facts, but islands of understanding
the new quantum regularities, which are gradually merged to continents
of knowledge allowing to turn the "wonders" of the new science into a
natural manifestation of the microwave world.

    $\diamondsuit $ It seems paradoxical that {\it the waves on lattices} 
(in the case of discrete space coordinate) can be hold as {\bf bound states 
on the smooth potential slope}: they cannot fly to infinity although no 
barriers hinder this, see Fig 7.13. On the lattices {\bf bound states can 
live even inside the "up side down" wells }, {\it above the states of the 
continuum spectrum}, see Fig 7.6. {\bf The tunneling through the up side 
down barriers} hanging from above forbidden zone is also possible on 
 lattice, see Fig.7.26.

The behavior of waves on lattices appears to be valuable like the model
for investigations of periodical structures and still hardly
understandable nonlocal interactions. So, the minimal nonlocal
perturbations allow one {\bf  to squeeze and broaden the spectral zones} in
 different configurational space regions, to {\bf produce the
inversion of spectrum} when, for example, the number of signs changes
("knots") of bound state wave functions is decreasing (!) with the
excitation energy, see Figs 7.21, 7.22.  This  {\it instructive exotic 
character of the discrete quantum mechanics } is supplemented by the 
generalized "Schr\"odinger equations" of higher (4th and more) order, 
which opens new possibilities for description of a simultaneous motion of 
waves of different kinds (with different frequencies) in the same 
direction.

The following remark will help understanding of the discrete problem
specificity. In some sense, the continuous problem spectrum corresponds to
only a lower half of the spectrum band $E< E_{sc}$ ($E_{sc}$
denotes the center of the spectral zone) in the problems with the discrete
coordinate (in the limit of vanishing lattice step, the second half
as if goes beyond the infinity $E_{sc} \rightarrow \infty$).
Generally,  {\bf discrete quantum mechanics is richer than the
ordinary continuous one and includes the latter as a
particular limiting case}.

$\diamondsuit $    What was said in the first sections of the book about the 
simplest one-dimensional systems can be extended to complicated 
many-dimensional objects which are conveniently described by the systems of 
coupled Schr\"odinger equations:  {\it multichannel formalism}.
Only there, {\it instead of scalar wave functions, potentials and
spectral weight factors, the corresponding vector- and
matrix-valued analogues are considered}.  It turns out, many
wonderful discoveries (first chapters) not only find
their own matrix-valued generalizations, but there opens a free range
for finding new phenomena having no one-channel analogues.
So the interactions were found which provide {\bf absolute
transparency at any energies of the continuous spectrum of
the systems even with barriers (!)}, see Fig.8.14,
 and without bound states at all,
see Fig.8.15.

$\diamondsuit $ Recently, {\bf new mechanisms of total resonance
{\it transparency} and absolute {\it nonpenetrability} were found,
which consist in the accumulation of waves in the closed channel and
their subsequent decay into the entrance channel, which results
in suppression of either reflected or transmitted waves}.
Analogous resonances were revealed in systems described by
differential equations of  higher ($\ge 4$) order.
It was found that  {\bf the same system has a paradoxical
ability to combine incompatible, one would think, features to be
transparent (100\%) and simultaneously completely reflective at one energy
point, and even retain the waves in a bound state -- confinement in the
continuum}. Such freaks of wave interference are possible at different
boundary conditions.  The acquaintance with such
peculiarities of quantum systems raises to a new level the understanding
of the possibilities of the quantum design.

$\diamondsuit $  You will see how, increasing one of the components
of the spectral weight vector (generalization of the factor
$c$), {\bf the waves of a multichannel system can be gathered in one place
 not only in the configurational space, but also in the channel
space}, see Fig. 8.3. In doing so, other states as if undergo the recoil to
the opposite side in both the spaces (the phenomenon of state separation).

 In multichannel systems, {\bf the spectrum can branch out}.
The algorithms of wave packet movement and spectral branch control
will be demonstrated.

$\diamondsuit $ A notion about interchannel oscillations of a two-component 
wave packet constructed from near spaced states of a narrow doublet will be 
introduced.  There is analogy with one-dimensional equal potential wells 
divided by barrier, through which  the wave packet  is tunneling hither and 
thither.  In the interchannel case the part of the separating barrier plays 
the weak coupling of channels and the motion is considered over the 
discrete channel variable.  This is  a significant addition to the 
algorithms of packet control.

In systems of several bodies the connected particles assist one
another to penetrate barriers. This effect does not reveal itself in
 the limiting cases (when the particles tunnel through the
barrier being independent or compressed to one point), but becomes 
distinctly apparent in the intermediate case of internal oscillating 
motions of coupled particles. There is also  symmetry violation 
of tunneling in the opposite directions through the nonsymmetrical barriers
in contrast with the one channel case.

All this enriches our quantum intuition and allows one to look into
mechanisms of the wave motion along labyrinths of the channels.
These secrets of quantum  multichannel "kitchen" were hidden before  
inside the black box of computer calculations.  

One can say that you keep in your hands a book of pictures-snapshots
from the reverse, invisible in the "old" approach, side of quantum 
mechanics, which {\bf supplements} known books of quantum pictures by 
Brandt and Dahmen \cite{BD}, Popov et al \cite{Popov} on the traditional 
theory of direct problem.

We supplied the text with exercises and comments which
will help a reader to learn contents of the book.

\newpage

Chapter 1 {\large \bf
Main notions about shifting positions of bound  state energy levels
(by  potential transformation preserving symmetry)
}

\vspace*{.5cm}
How to deform the potential in order to change in the desired way the 
disposition of separate energy levels $E_{\nu }$, the most important 
spectral characteristics?  Such elementary transformations reveal  
connections between observables and the related forces acting in quantum 
systems.  Here are explained the simple qualitative rules, discovered by 
us, of lowering and raising an {\it individual} energy level $E_{\nu }$. It 
would be difficult to guess how
to do it within the traditional approach of the direct problem.  In fact,
this possibility has not hitherto been mentioned  in most up-to-date
manuals; though the acquaintance with  these samples of `control' of quantum
waves is the best way of comprehending the microworld.

The qualitative essence of the majority of quantum system
transformation algorithms, to be discussed here,
can be distinctly revealed by the simplest models. So we will
start with the infinite rectangular potential well of width $\pi $ and
look how its relief changes under variations of a selected energy level
value $E_{\nu }$.  The potential perturbations are especially visual
against the background of the flat bottom of the chosen initial well.

At first, it is better to restrict ourselves to the potential
transformations which {\it do not violate the symmetry} of the initial
potential $\stackrel{\circ}{V}(x)=0$  with respect to its central point
$V(x)=V(-x)$. In this case, only the {\it the
purely discrete energy levels} $E_{n}$ form {\it the complete set of
spectral parameters} which uniquely determine the potential shape.
Really, for determination of a symmetric potential it is {\it
sufficient to have one its half} because another one is the same. So,
although for the arbitrary infinite deep potential well all the energy
levels are only {\it the half} of spectral data, they are enough
to determine a symmetric system exactly. In the unsymmetric
case $V(x) \ne V(-x)$, there will also be needed  {\it spectral weights} of 
each bound state, but this will be discussed in the next sections of this 
book.

 How should  the potential bottom be deformed to shift {\bf the only} 
chosen energy level (spectral "brick") while keeping the other ones at 
their previous positions? It is very important to understand just these 
{\bf elementary} transformations to have a notion about the methods of 
constructing arbitrary systems with the given properties.
Let us at first move the lowest energy level with the most simple wave
function.  Here the following question is helpful. In which place of the
well is the ground state most sensitive to the potential perturbations?
Of course, where the probability of finding the particle has a
maximal absolute value.  For  the ground state of the infinite rectangular
well it is its central part (see the unperturbed wave function
$\stackrel{\circ}{\Psi}_{1} $ which has the only bump of $\sin (kx)$, in
Fig.  1.1a,b). Hence, for instance, to {\bf shift down} the first level
it is necessary to shift down the bottom of the potential just
at the center, as is shown in Fig. 1.1a,c \cite{PoshTrub}.  The nearer to
the ends of the wave function bump (to its knots at the walls of the initial
well) the weaker is the sensitivity. At the boundary points (at the
walls) the function vanishes, it completely ceases to respond
to potential perturbations there.

\fbox{Place of Fig. 1.1}

However, the potential well perturbation is not sufficient for our 
task.  If the potential perturbation is purely attractive, it will shift 
down {\bf all} the levels; but our goal is to consider the {\it simplest, 
elementary} transformations changing  only {\it one} spectral parameter and 
leaving all remaining parameters at their previous places.

It is clear that we have to add a {\it compensating
repulsion}, the potential hills at the walls of the initial well, see
Fig.1.1a,c.  They will weakly influence  the ground state, because
inside the region of their action the wave function is small (near the
knots at the walls).
  The different situation is for the excited states :
 there the absolute values  of their wave functions
are somewhat bigger than those of the shifted ground state and vice versa
at the center. In the process of
lowering of the ground state $E_{1}$ the central well in $\Delta V(x)$ is
deepening, and the barriers from both sides become higher, see Fig.  1.1á.

This picture is very instructive because it demonstrates the qualitative
features of the {\it universal, elementary} transformation of
$\stackrel{\circ}{V}$ and $\stackrel{\circ}{\Psi}_{1} $ on the interval of
\underline{one bump}. This allows one to understand the rule of
transformation of potentials and waves for \underline{any} shifted 
states with \underline{several bumps} and in \underline{arbitrary} 
potentials, see the following figures. It should be emphasized that 
although all levels $E_{n\ne1}$, except the chosen one, $E_{1}$ are 
conserved, the wave functions of {\it all} states $\Psi_{n}(x)$ undergo 
distortions.

 The exact  form of the potential perturbation is determined by the
formalism of supersymmetry in quantum mechanics  SUSYQ (\ref{v2}). For the
concrete case of the initial rectangular well the transformations have the
following   exact  expression (the interval $[-\pi/2,\pi/2]$)

\begin{eqnarray}
V(x)=\stackrel{\circ}{V}(x) \nonumber \\
+2 t \frac{d}{dx} \{\sin(\sqrt{1+t} x)
\cos(x)/[\sin(\sqrt{1+t} x) \sin(x) +
\sqrt{1+t} \cos(\sqrt{1+t} x) \cos(x)]\}
\label{vE1shift}
\end{eqnarray}

with  the expression for the  normalized eigenfunction at the new
eigenvalue $E = 1+t$ is (see the general case see in
(\ref{psit}) :
\begin{eqnarray} \Psi (x,1+t)=\cos(\sqrt{1+t} a)/
\{\sin(\sqrt{1+t} x) \sin(x) \nonumber \\
+ \sqrt{1+t} \cos(\sqrt{1+t} x)
\cos(x)] \}.  \label{psiE1shift}
\end{eqnarray}

We shall postpone the derivation of analogous expressions in the general
case for any potentials  till chapter 3. For the present,  it is important 
to know that  the initial $\stackrel{\circ}{V}$ and 
$\stackrel{\circ}{\Psi }$  serve as the building material 
 for the new transformed potentials  $V$ and functions $\Psi 
$(in our case, the usual sines and their derivatives).

Pay attention to the simplicity of exact expression
(\ref{vE1shift}).  It is wonderful, because one could hardly suppose that
after many tens of years since the creation of quantum mechanics there
would appear a possibility of achieving, with so elementary tools, the 
result which seemed impossible. Really, this is analogous to a virtuous 
surgery on transplantation of a level to a new place
without any "injuring" the positions  of all other levels.
So, it proves to be feasible  to satisfy the infinite(!) number of
additional conditions. To facilitate the believe in
that, let us remind that all the eigenstates are orthogonal and, in some
sense, absolutely unlike one another. This circumstance plays a
principal role in the IP formalism.

It is of great importance to inform as soon as possible  broad
physicist circles, particularly the lecturers on quantum mechanics,
about the revelation of such
constituent elements of fundamental spectral transformations.

  The first our task is to concentrate one's attention on the qualitative
understanding, "from the first principles", of quantum design. 
  Further, it will make easier to master the mathematical
formalism: after clarifying the physical essence it will be easier to
follow the details of the corresponding mathematical expressions.
  It is not sufficient to know the mathematics of the inverse problem, 
its equations, formulae of exact models, etc. It is necessary to have a 
physical intuition allowing one to heuristically (without computers and 
analytical mathematical apparatus) foresee the algorithms of how to 
transform a potential to get the desired properties of the considered 
system.

  The significance of energy level position can be illustrated,
for example, by the fact that "a quantum of the light exciting the
chlorophyll molecule transfers one electron on a  special level.
And in a fraction of second it returns into the initial state {\bf
triggering the mechanism of any life activity on the earth}.
Though only a few organisms (plants and some bacteria) contain chlorophyll
providing the photosynthesis and transforming the energy of light, but
owing to them it becomes available for all others including the mankind"
(P.Raven, R.Evert, S.Eickhorn, Modern botany).  So, {\bf our own
existence depends on the position of  one quantum energy level}.  Although
we still cannot change in practice the spectrum of chlorophyll in the
desired way, but it is important to acquire the understanding of 
the principle of the relation  $\Delta V(x)\leftrightarrow \Delta E _{\nu} 
$).  At the same time,  the (nano-) technology which can be used for 
creation of the desired effective potentials in electronics (superlattices, 
 quantum optics with the variable refraction  coefficients, etc.) is 
 intensively developing.  However, in the theory, the possibility of
influencing  the spectral parameters provides a 
deeper insight into  the internal logic of constitution and behavior of any 
 quantum system, even if we do not intend to change real objects.

It is wonderful that all this  can be seen using the {\bf exactly
solvable models} which appeared to be  so numerous (even complete sets !)
that they can in principle to approximate arbitrary cases. As in the first
example all will be clear  in this chapter on model pictures without 
formulae.

So, it is natural to suppose that for  {\bf shifting up} the first
energy level $E_{1}$, the bottom of the potential must be lifted in the
center, as is shown in Fig.1.2.
\cite{PoshTrub}.

 However, as in the previous case, if the potential perturbation is
purely repulsive, it  shifts  \underline{all} the levels  of the
spectrum. Our goal here is to keep other levels (except the chosen
one) at their  initial places.
Let us repeat the following: for a deeper understanding of the spectral
engineering one should learn most elementary transformations.
Hence, it is necessary to create {\it compensating
attraction}, i.e. potential wells must be introduced at the places where
the sensitivity of the ground state to the potential perturbations is
weakened : in the vicinity of knots near the walls of the initial well, see
Fig.1.2a,b.  Equation (\ref{vE1shift}) gives just such a special form
of the barriers and wells to shift the energy $E_{1}$  of the bound state
up while keeping the levels of exited states fixed.  For them
the potential barrier is completely compensated by additional wells as in
Fig.1.2a. In particular, it is evident even without formulae that these
wells pool down the first exited state $|\Psi_{2}(x)|$, having its two
bumps apart from the center, more strongly than the $\Psi_{1}(x)$
whose absolute value is smaller near the walls.

As the ground state level $E_{1}$ is lifted to its neighbor $E_{2}$, the
central barrier  in Fig.1.2b increases and the wells at the edges become
deeper. The perturbations shown in Fig.1.2 ,b keep the second level
from shifting due to a mutual balance between attraction and repulsion.

  The profiles of perturbations  $\Delta V(x)$ in Fig.1.2a,b are like
the upside-down potential curves in Fig.1.1 ,c. for the case of lowering
the ground state. The same "inversion" appears also for  upward-downward
shifts of other states and in the cases of different initial
potentials.

As the energy levels come close together, the absolute values of the
corresponding wave functions  $|\Psi_{1}(x)|$ and $|\Psi_{2}(x)|$ become
more alike. It is shown in Fig.1.2c how the wave function of the ground
state is changed: it gradually splits into two hills in separate
potential wells. The value of $|\Psi_{1}(x)|$ in the central part
decreases (barrier forces it out); and in order to push up 
 this two-humped ground state, the two-humped barrier is
needed, see the minimum beginning to show in the center of the barrier in
Fig.1.2b.  Several curves in Fig.1.2b,á, corresponding to the evolution of
the potential and the wave functions allow one to imagine qualitatively
the whole continuous manifold of their intermediate forms.
All the curves shown belong to the class of the exactly solvable models, see
section 3.3.4. The unexpected "annihilation" of states in the limit
of total coincidence (degeneration) of levels is shown in Fig.5.1. This
incompatibility at one energy of several one-dimensional quantum bound 
states is the bright characteristic of eigenfunction properties.

Now when the qualitative explanations of $\Psi , V$ transformations are
found they seem so natural, but it was before difficult to guess
them because it seemed that there was no hope at all for such a gift of
quantum theory.  The  generalization of these
rules to all the states of the discrete spectrum and arbitrary symmetric
potentials, as will be shown further, was even more unexpected.  
The  revealing of these rules suddenly dawned upon us.  Before this, it 
even did not come to our mind that behind rather exotic bends of $\Delta 
V(å)$, appearing on the PC screen as a result of millions of numerical 
operations, there could be hidden clear, very significant and simple 
physics.  This is an example of the everlasting magic of comprehending the 
truth:  turning the inconceivable into the evidence.

 Let us compare the forms of
the potential curves  $\Delta V_{1}(x)$ and $\Delta V_{2}(x)$, shifting
upward and downward the ground state in Figs 1.1, 1.2 and the exited one in
Figs 1.3, 1.4. There is a
simple regularity in repeating influence of force on  separate bumps
of the corresponding functions.  Now it is easy to explain the shape of
potential perturbation for the upward (downward) shift, for example, of only
the {\bf second} level in the infinite rectangular well. The corresponding
unperturbed wave function represents one period of $\sin(kx)$ with two
bumps and one knot in the center.  So, for shifting downward (upward) the
second level there must be in the perturbation potential, already two
minima (maxima) of attraction (repulsion) at the places where
the $|\Psi_{2}(x)|$ has maxima as shown in Fig.1.3, 1.4.
This can be seen from Eqs (\ref{psit}) and (\ref{v2}).  Three
barriers (wells) near three {\it knots} of $|\Psi_{2}(x)|$ keep  other
levels from moving (down-) upward. For the inverse problem such a control 
is a simple procedure because (as mentioned before) in that approach the 
level positions are the entrance parameters.  Here the absolute precision 
of the level control is possible again due to the mutual orthogonality of 
all the states of the spectrum, the difference of their bumps and their 
displacement.

It is remarkable that all this is performed with only
the local transformations of the interaction $\Delta V(x)$.
In the direct problem there were already known  shifts of separate 
energy levels, but it was done by \underline{nonlocal} (separable) 
potentials often considered as less convenient for description of 
quantum systems from a physical point of view.

 In the process of shifting the levels $E_{2}$ to
$E_{1}$ ($E_{3}$), the form of the absolute value $|\Psi_{2}(x)|$ bears
a greater resemblance to the $|\Psi_{1}(x)|$ ($|\Psi_{3}(x)|$).
This will be more clearly shown further  when considering a very
interesting transition to the limit of the complete coincidence of $E_{2}$
with $E_{1}$ ($E_{3}$) and their {\it effective annihilation}, see Figs 
5.1, 5.6.

The above considerations completely apply to the case of wells
with nonvertical walls. See for example, the deformations of the
harmonic oscillator potential in Fig.1.5 -e \cite{PoshTrub}.
Pay attention to the analogy of potential deformations there and in Fig.1.2.
 In Fig.1.5c-e the shapes of $V,\Delta V$, lifting the ground
states in the rectangular and oscillator wells are compared.

We can detach the lowest state with zero energy $E=0$  
from the free motion continuous spectrum  and  transform it into the 
bound state of the soliton-like reflectionless potential well, see Fig.1.6.
For this purpose, we shall use "the intuition" we have 
acquired in shifting down the state of discrete spectrum. Though the 
initial systems seem to differ very much from one another, they become 
alike in the limiting case when the walls of the rectangular well go to  
$\pm \infty $:  transition to the infinite wide well. The form of the 
potential perturbation $V_{1}$ (the soliton-like well in Fig.1.6 ) we 
search for, can be explained in the following way.  It occurs as a result 
of lowering the "scattering" state with the energy $E=0$. The wave length
of this state is infinite and the unperturbed wave function 
represents a horizontal line corresponding to the single bump of sine with 
the infinitely small frequency. Shifting down the state with {\it one} bump 
must be performed according to the rules with only {\it one} well. The 
compensating repulsive barriers are "not seen" 
because the knots of a bump with zero wave number move to $\pm \infty $ and 
disappear along with these barriers.  When  detaching  the lowest state 
from the continuum, it is possible to pull down it to the arbitrary 
depth in the negative energy region.  In the case of the discrete spectrum,
 the shift of one level leaves all other states unaltered. Here are 
conserved the characteristics of the continuous spectrum, the 
values of reflection coefficient at different energies.  So, the property 
of free waves having no reflection is inherent in the soliton-like wells.  
After the detachment from the continuum of one bound state, there will 
remain, as the lowest scattering state $E=0$,  a
wave with one knot and two infinitely long bumps.  For tearing also this 
state into the discrete spectrum, there will be needed, according to 
the same logic, two wells and one barrier of potential perturbation 
$V_{2}$, see Fig.1.6b.  Meanwhile, the new wave function with $E=0$ gets 
the second knot.

If one takes, as an initial potential, a hill which
gives considerable reflection, then creating a single bound state by
symmetric potential perturbation  [\cite{Suk}, 1985] will give a well with
barriers at both sides with the same reflection as for the initial
barrier, see Fig.1.7.  The same can be achieved with a well being aside
of the potential hill, so one obtains a non-symmetrical $V(x)$.
Another example of shifting down a level  represents a  $\delta  $-well
with the only bound state, see Fig. 1.8, when compensating potential
barriers are not required.

As an additional confirmation of the rules formulated above, 
the transformations of the oscillator well are shown in Fig.1.9a,b:  (a)
the 20th and (b) the 30th level are shifted upward. Here  $\Delta V(x)$ has
(a) 20 and (b) (30) hills and  21 (31) wells.  It is remarkable that 
lower states  {\it practically are not changed}. This is because of a 
strong inertia of low-lying states with respect to frequent oscillations of 
the perturbative potential. Their influences  almost cancel each other when 
averaging. Only the high states are significantly changed near the shifted  
20th (30th) level.  There is a conformity between the  wave functions  and 
the perturbation $\Delta V(x)$ in that energy region, due to their 
commensurable frequencies (some kind of "resonance").

The shifts of levels $E_{n}$ can be performed not only over the real
 energy scale, but the chosen levels $E_{n}$ can be moved into the complex
energy plane.  The resulting unusual non-Gamov decaying states will be 
considered in section \ref{non-gamov}.

Many aspects of quantum design have opened for us  in the process of the
 intensive work on this book. We think that, thereby, we have to some 
extent met the requirements of the nontedious writing declared by L.  
Levitsky in his diary which we have recently found by chance :  "The sense 
 of book writing  consists in that it first of all opens something to its 
 author. I am sure that good books differ from the bad ones in that the 
 authors of the first ones, beginning to write them, have cleared up 
 something for themselves, while authors of the others have beforehand 
 known everything". So, in particular, while writing this book 
  there appeared an 
 interesting hypothesis.  It was later confirmed by exact calculations.  It
 appeared that using the above-mentioned potential perturbations like
  (\ref{v2}), but continued periodically, we can {\bf control the position
 of allowed and forbidden zones}  of band spectra, see 
section \ref{period}.  The zones can be moved apart or drawn together up to  
their merging (disappearance of lacuna dividing them).  Later we shall show 
 a set of the corresponding examples.

\newpage

FIGURE CAPTIONS

Fig.1.1 a) Potential
perturbation of the infinite rectangular well which causes lowering (shown
by the arrow $\Delta E_{1}$) of only one ground state level $
 \stackrel{\circ}{E}_{1}=1 \to E_{1}=-4$. The thick black arrow directed
 down shows the well (dashed-dotted line $\Delta V(x)$) that influences
the most sensitive central part of the wave function and pulls the level
down. Black arrows directed up near the walls of initial well, i.e. 
near the knots of $\Psi_{1}(x)$ where it is the least sensitive to 
potential variations, point  to the barriers which keep fixed all other 
levels $E_{i \ne 1}$. The thin dashed lines show the transformations of the
wave functions $\stackrel{\circ}{\Psi}_{1,2} \to \Psi_{1,2}$.  All this 
demonstrates the qualitative features of {\it universal, elementary} 
transformation for a {\it single} bump. This will allow one  to understand 
the rule of transformation for arbitrary states with {\it many} bumps and 
for {\it arbitrary} potentials.

b) Evolution of the function $\Psi_{1}$ for $\Delta E = -1, -3$

c) Evolution of the potential for $\Delta E = -1, -2, -3$.

Do not confuse b,c) with real motions in some quantum system: these are
  variations of $\Psi_{1}, V$ during gradual transitions from one system
to another.

 Fig.1.2
a) Deformations (dashed-dotted line) of the rectangular well ($V(x)=0$)
raising the ground state level $\stackrel{\circ}{E}_{1}$ to $E_{2}$.
Pay attention to the form of transformed potentials $V(x)$ in this and
previous figures: they are qualitatively similar up to the change of the 
sign.  The barrier in $\Delta V$ in the sensitive center of the wave 
function $\Psi_{1}$ pushes $E_{1}$ up (shown by thick black arrow). There 
 are wells from both sides pointed by thick black arrows down, where 
$\Psi_{1}$ is less sensitive. They weakly influence on $E_{1}$, but 
neutralize the barrier influence on all other levels $E_{i>1}$ keeping them 
fixed. The dashed lines show the energy levels and the deformed wave 
functions $\Psi_{1}$, $\Psi_{2}$. This is another example of level control 
for state with {\bf only one bump}. It will allow to clarify the general 
case of shifts of arbitrary levels in arbitrary potentials.

b,c) Evolution of potential  $V(x)$ and function $\Psi_{1}$ for
 $\Delta E = 1,2,2.5$. Pay attention to the deformation of $\Psi_{1}$.
Its shape approaches to absolute value of $\Psi_{2}$. This is due to 
 a tendency to lowering the central barrier in the middle in Fig.1.2 .

Fig.4.1
Demonstration of approximation of the (a) linear, (b) oscillator and (c)
 rectangular infinite potential wells by the reflectionless wells of
finite depth when the lower parts of their discrete spectra  coincide 
(eight lower bound state energy levels)  \cite{Schon}.

Fig.4.2  Stepwise well of finite depth with 8 equidistant bound
states energy levels approximates the lower part of the oscillator well in 
the energy region of spectral coincidence (T.Stroh) although in the upper 
parts their spectra are quite different (discrete and continuous).

 Fig.6.22 a)  Shift of the upper  boundary of the second 
allowed zone  for "Dirac comb", periodic  $\delta $-barriers.  This 
boundary coincides with the second energy level of the auxiliary 
rectangular well of widths equal to period and with the unpenetrable walls  
instead of $\delta $-peaks.  Shift of this level up by $\Delta V(x) $ in 
the partial well will shift the whole corresponding zone in periodically 
continued potential perturbation $\Delta V(x) $. The level pools its zone 
  while the lower boundary of the neighbor zone becoming wider comes down 
to meet the zone below. This motion leads to disappearance of the gap 
between the second and the third zones at $\Delta E=1$. So the second level 
of the auxiliary well becomes also the lower boundary of the upper zone. 
Then the zones go apart $\delta E=2,3,3.9$, but the shifted level is 
"sticked" to the other zone. It squeezes this zone till annihilation 
because the upper boundary $E=9$ is fixed according to the algorithm of 
shifting the only one energy level.  b) Shift down of the upper boundary 
$E=4$ of the allowed zone in Dirac comb. This boundary pushes its zone down 
until its lower boundary  reaches the lower allowed zone. Then its 
fixed upper boundary $E=1$ becomes also the lower boundary of the lowering 
zone. After the merging, the lower zone is teared off the upper zone and 
goes down and the upper zone is squeezed by its boundaries: the fixed  
$E=1$ and the lowering second auxiliary energy level.  c) Shift up of the 
lower boundary of the second allowed zone of the $\delta $ wells comb (of 
the ground state level of the auxiliary eigenvalue problem on the period). 
This boundary pushes its zone until its upper boundary reaches the position 
of the second fixed energy level  $E=4$ and the lacuna between zones 
disappeared. After that the zone squeezing begins  as in the case b).  Then 
the upper zone splits from the lower one and goes up restoring the 
 disappeared lacuna.  The annihilation of the auxiliary energy levels leads 
to disappearance of all allowed zones. 

 Fig.1.6  a) The lowest state of the continuous spectrum of
the free motion with $E=0$ is splitted down and becomes  a bound state
 $E_{1} < 0$. The perturbation  $\Delta V(x)$ shown  by the arrow is a 
soliton-like well $V(x)$. It has no barriers unlike the parturbation in 
Fig.1.1. It is because the "knots" of the $E=0$ state disappear at $\pm 
\infty $ and because it must not change the zero reflection of the initial 
free waves, although all wave functions undergo transformation.  For 
example, the new state with $E=0$, which does not increase at infinity, has 
a knot and horisontal asymptotics.  b)  Two bound state levels are teared 
from the continuous spectrum (arrows $\Delta E_{1}, \Delta E_{2}$).

Fig.2.1 The (a,b) transformation  of the infinite rectangular well
 $\stackrel{\circ}{V}\rightarrow V$ and eigenfunctions 
$\stackrel{\circ}{\Psi}_{1}(x)$ (a,d),
 $\stackrel{\circ}{\Psi}_{2}(x)$ (a,c) by increasing SWF
 $\stackrel{\circ}{c}_{1}\rightarrow c_{1}$, the derivative  $\Psi'_{1}(x=0)$
at the left wall.  The scales of the functions  $\Psi_{1},\,\Psi_{2}$  (a) 
are shifted up to the corresponding energy levels  $E_{1},\,E_{2}$.  In 
(b,c,d) the evolution of the potential and functions with increasing 
 $c_{1}$ 2, 5, 10, 20 times is demonstrated. Meanwhile {\bf all !} energy 
 levels $E_{n}$ and all SWF, except  one $c_{n \neq 1}$, remain unchanged, 
as is seen for  $\Psi_{2}$ (a,c). The parameter SWF $c_{1}$ controls the 
localization of the wave function  $\Psi_{1}(x)$ (a,d,e) in space: by 
increasing  $c_{1}$ the ground state is pressed to the left potential wall, 
as is shown by arrows in (a,d). This is performed by the potential barrier 
in (a,b) on the right which shifts the function $\Psi_{1}(x)$ to the left, 
and the potential well on the left which simultaneously  compensates the 
influence of the barrier on  {\it  the energy levels}  and keeps them all 
at the same places. All wave functions, except the ground state, undergo 
some recoil in the opposite direction which is demonstrated by   
$\Psi_{2}(x)$ (a,c). So there is separation of the bound state from others. 
Compare with Fig.2.2, where the SWF $c_{2}$ was changed.  (f) 
Transformation of the oscillator potential (barrier+well) by shift of the 
ground state to the right.

Fig.2.2  Deformation of the rectangular potential and functions by 
increasing SWF $c_{2}$, the derivative $\Psi'_{2}$ at the left edge of the 
well.  (b,c,d) evolution  of the  potential and functions  with increase in
SWF $c_{2}$  2,3,6 times. The scales of the functions  
$\Psi_{2},\,\Psi_{3}$ (a) are shifted up to the corresponding energy 
levels   $E_{2}\,,E_{3}$.  The position of all  energy levels  $E_{\lambda 
}$ and SWF  $á_{\lambda \neq 2}$ (b,d) is not changed. The potential 
perturbations $\Delta V(x)$ (a,c) consist of two blocks "well-barrier", 
each for one of two bumps of $\Psi_{2}$. Compare with Fig.2.1. The 
 influence of these blocks reveals in shifting the maxima of each 
bump in $\Psi_{2}$ to the left. The relative values of these bumps also 
increase to the left due to a smooth connection of $\Psi_{2}$  at knots 
with violated symmetry of derivatives $\Psi'_{2}$ at knots.  Pay attention 
to that the central knot of $\Psi_{2}$ remains at the same place.

Fig.2.5 a)
 "Pressing out"  the ground state of the finite rectangular well
$\stackrel{\circ}{V}$ by decreasing  SWF  $c_{1}=\Psi'_{1}(x=0)$.
This shift is performed by a universal potential block barrier+well $V(x)$.
In this case, the soliton-like reflectionless well  and an additional 
potential peak at the edge of the initial well compensating the smoothing 
of the potential step, provide the invariance of all other spectral 
parameters of discrete and continuous states.  The bound state wave 
function  $\Psi_{1}$  concentrates in the carrier well shifted to the right 
where are the conditions for constructive interference of multiply 
 reflected waves in a soliton like well at $E=E_{1}$.  At the place of the 
initial rectangular well  $\Psi_{1}$ is negligibly  small due to a
destructive interference of waves multiply reflected from the walls of the 
narrowed well. Wave functions of other states are somewhat recoiled to 
the origin. For instance,  $\Psi_{2}$ 
 in contrast  with $\Psi_{1}$ is concentrated near  $x=0$,  as if it 
becomes the ground state in the left well and its second bump inside the 
right well (it is shown with 1000 times enlargement) is negligibly small 
because of destructive interference.

b) Comparison of the initial and  transformed functions
$\stackrel{\circ}{\Psi}_{2}$, $\Psi_{2}$  of the second bound state.
The curved  arrow shows how the probability distribution undergoes recoil
to the origin without changing  SWF $c_{2}=\Psi'_{2}(x=0)$. The second 
bump of the function  $\stackrel{\circ}{\Psi}_{2}$ is almost pumped into 
the first one of the transformed wave  $\Psi_{2}$ and the knot  
disappears in the limit  $c_{1} \rightarrow 0$.

Fig.5.1
The graduate deformation of the infinite rectangular well and functions of 
bound states when  $E_{2}$ approaches $E_{3}$,  and the effective 
annihilation (they are pressed into the potential walls and disappear 
behind the infinite potential barriers) of the degenerated 
states, compare with Fig.1.4.  The states above the degenerated ones 
gradually lose the first and the last bumps with the corresponding knots. 
Inside the new potential these states have numbers lowered by 2.

Fig.5.7 Symmetrical deformation (b,c,d)  of the oscillator 
 potential (a) $\stackrel{\circ}{V}(x)=x^{2}$  during coming together 
 of bound state levels $E_{3}$,  $E_{4}$ (b,c) in the limit of their 
 degeneration (d).  First 8 wave functions are shown. The waves  
 $\Psi_{3}(x)$ and $\Psi_{4}(x)$, each teared  into two parts, are 
concentrated inside special carrier wells and moved  in the limit of total 
degeneration to $\pm \infty$ (annihilation).  The wave functions of 
the states below the degenerated doublet are not radically changed. 
However, $\Psi_{5}(x)$ and others  $\Psi_{m}(x), m>5$ have decreasing first 
and last bumps  (b -- d) disappearing in the limit, see  $\Psi_{8}(x)$
 ($b \to c \to d$), compare with  Fig.5.3.

Fig.6.7. Demonstration of the control of resonance widths.
  The delay time of scattered waves inside the potential in which the first
bump of the lowest quasibound state is shifted inside the soliton-like well
through the potential barrier to its outer edge to make its decay easier.
Compare with Fig.2.5. As a result, the width of the first resonance becomes 
broader than for the second one.

Fig.2.14  Creation of a bound state  at the negative energy (a)
and in the continuous spectrum  (b-e) \cite{StilHer}

Fig.6.13. On the right side  ($x \ge 0$) the potential
 $V_{BSEC}(x)$ is shown which transforms the free wave (sinus) into BSEC
$\Psi _{BSEC}(x)$ confined at the origin. The axes for $\Psi _{BSEC}(x)$ and
$V_{BSEC}(x)$ are shifted to avoid their overlap. Pay attention to coincidence
of bumps and knots  of the BSEC  wave function with blocks "well-barrier" 
and even knots of the perturbation potential. Just this correlation 
provides the confinement of waves near  the origin of the pinned out 
point $E_{BSEC}=1$ of the energy scale. At energy points different from 
$E_{BSEC}$ the physical solutions are not decreasing in asymptotics. On the 
whole axis $x$ there is a total reflection of waves (symbolical arrows 
on the left side of Fig.) by the potential $V_{r}$ continued to the left 
$x$ half-axis as zero and coinciding on the right half-axis with 
$V_{BSEC}(x)$. Only one solution at $E=E_{'''}=1; \enskip c=1$ is 
decreasing when  $x \to \infty$ (the other independent one is nonphysical 
and increases).

Fig.6.14.  The absolute value of the reflection coefficient $|R(E)|$
for the potential with BSEC on the half axis  $0\le x < \infty $  at energy
$E_{BSEC}=10$ with resonance reflection (100 \% at this point) for waves on
the whole axis. With increasing SWF of BSEC $c$ the width of the resonance
of  $|R(E)|$ near $E_{BSEC}=10$ becomes greater. For  $c\rightarrow 0$ this
width converges to zero: the dashed line corresponds to this limiting $R(E)$.

Fig.7.13  The bound state wave  functions on the linear slopes
 ( , $b$, $c$): $V(\alpha )~=~\alpha ;  \alpha /2; \alpha /4 c$ of the discrete
variable  $\alpha $  are the Bessel functions $J_{\alpha }$ of index equal 
to whole numbers. Do not confuse with the usual argument of Bessel 
 functions $kr$ which is fixed in our case (!).  The values of the 
functions depending on the discrete variable $\alpha $ are connected by the
dashed line to make the function easier to imagine.  Here the parameter  
$z$ does not play the part of a coordinate, but regulates the steepness of 
the potential slope.  The less steep is the slope, the more dense becomes 
the equidistant bound state spectrum E$_{n}$.  The functions of all these 
states for the fixed $V(\alpha )$ have  the same form, but with the 
argument shifted by the whole number. For the nonwhole number $\alpha $, 
but with the whole value of the lattice step, the Bessel functions are 
nonphysical states with increasing absolute values in the direction to the 
left.

Fig.7.6.  Bound state in the up side down well above the continuous
spectrum of the allowed zone has the characteristic change of the sign. 
Nevertheless the enveloping  function of the upper bound state is similar 
to the function of the ordinary bound state  in a usual well (in our case 
of the ground state).  As in the case of continuous variables,  there is 
enough arbitrary small deepening  of the upper boundary of the allowed zone 
into the upper forbidden zone for the appearance of at least one bound 
state.  Compare with the form of the wave functions in the infinite 
rectangular well on the lattice, see Fig.7.21.

  Fig.7.26  The local potential  $V(x)$ bends the allowed zone
in the plane (E,n) without changing its width. Compare with the influence of
the minimal nonlocal part of the potential which controls the width of the
zone, as is shown in Fig.7.25. The discrete values of the functions are
connected for clarity by the solid lines.

Fig.7.25   ) Potential $u(n)$, which determines the mutual 
influence of the neighbor points (minimal nonlocality), controls the change 
of the width of the allowed zone  (b) at different points n.

Fig.8.14 The transparent interaction matrix with one bound state
at the energy $E_{b}=-0.5$  with SWFs  $M_{1}=1,  M_{2}=0.001$ and different
thresholds $\epsilon_{1}=0,\> \epsilon_{2}=1$, at which the channels become
open. Besides the soliton-like well  $V_{11}(x)$ on the right there is a
  part of $V_{\alpha \beta }(x)$ on the left (see Fig. 8.15) which strongly
mixes the components of the channel waves. The barrier in  $V_{11}(x)$
shown by the solid line  reflects waves which are suppressed by the waves 
"decaying" from the "quasibound state" concentrated inside the well 
$V_{22}(x)+\epsilon_{2}$" shown by the dashed line. At small energies, the 
waves which have to tunnel through the barrier in $V_{11}(x)$ go round 
it through the second channel using the channel coupling 

$V_{12}(x)=V_{21}(x)$ (dotted-line) The same 
as in (a) only the wave function components  $\Psi_{1}; \,\, \Psi_{2}$ of 
the bound state are shown.  The small component  $\Psi_{2}$ is shown 
 800 times enlarged.  á,b) The violation of conservation of 
the partial wave fluxes  $J_{\alpha }$ occurs only inside the left part of 
the interaction matrix and outside the fluxes return to their previous 
unperturbed values.  The case is shown (d) of the incident wave only ??? in 
the first channel which is transferred into the second channel, but then 
completely (!) returns to the first one even with the energy $E>\epsilon 
_{2}$ when both the channels are open.

Fig.8.15  The two-channel interaction matrix with nonequal thresholds
$\epsilon_{2} =1$) completely transparent at all energies of the continuous
spectrum and having no bound states. So, this $V_{\alpha \beta} (x)$ has 
no analogue in the one-channel case where there are no reflectionless 
potentials without bound states.  The remarkable peculiarity of this 
interaction matrix is the barrier in  $V_{11}(x)$ which does not violate
the transparency of the system although intensively reflects waves. 
However, the reflected waves are suppressed by the waves coming from the 
second channel due to the coupling $V_{12}$) and going backward with the 
same amplitude and the opposite phase as the reflected waves (destructive 
interference).  It is interesting that this interaction matrix is a 
constituent  part of the matrix with the bound state shown in Fig.8.15 .

\end{document}